\documentstyle[12pt,epsf]{article}
\input epsf
\begin{document}

\begin{titlepage}
\thispagestyle{empty}

\begin{center}
$\vphantom{begin}$

\vspace{1.5cm}
{\LARGE Vacuum structure, spectrum of excitations and low-energy phenomenology
in chiral preon-subpreon model of elementary particles.}

\vspace{2.2cm}

{\Large\sl O.E.Evnin}

\vspace{1cm}

{\large\sl Rostov State University.

Zorge avenue, 5, Rostov-on-Don, 344014, Russia.

\vspace{5pt}
E-mail: oevnin@uic.rnd.runnet.ru}

\vspace{4cm}
{\large\bf Abstract}
\end{center}

\vspace{0.5cm}
Inner and empirically consistent model of elementary particles,
including two matter structural levels beyond the quark one is built.
Excitations spectra, masses and interactions are analysed using the
phenomenological notion of non-pertubative vacuum condensate.
Essential low-energy predictions of developed concepts are classified.

Effective gauge $U(1)\times U(1)\times SU(2)$-theory of quark-lepton
excitations behavior based on the performed analysis of preon-subpreon
phenomenology is consistently built. The ability of its expansion with
fermions and scalar leptoquark coupling is also considered.
Shown that the coupling
constants generation hierarchy is the same as generation hierarchy of quark masses.

Using the built theory cross-sections of $e^+d\to e^+d$ and
$e^+d\to u\bar\nu_e$ processes are calculated. The obtained resonance peak
is proposed to be a possible explanation of deviating from Standard Model
predictions discovered in DESY in the beginning of 1997 year.

\end{titlepage}
\newpage

\section*{Introduction}

Standard Model (SM) which nowadays is thought to be a proper tool for elementary
particles researches containes two clearly distinguishable concepts of
physical vacuum. We mean Higgs vacuum of scalar particles, linear by field
and having classical origin, and QCD vacuum provided with non-pertubative
quantum fluctuations of quark-gluon fields. These different
notions straightforwardly produce the ideas of spontaneously broken and
hidden (confined) symmetry. Such a state of affair within the theory
makes to think about evolution of vacuum concept in future physics.
Two usual answers on the question are given by the two main expansions of SM.

Grand Unification Theories (GUTs) suppose Higgs mechanism and massive
vector bosons are both fundamental. For gauge structure of the theory it means that
we need to build a solid simple gauge group, including all known interactions.
When the symmetry, corresponding to it, is broken we come to the
observable manifold of physical phenomena.

Preon models take strongly into account, that only QCD vacuum, but not the
Higgs one is certainly experimentally observable. Consequently, it is hidden
symmetry concept which is of a big value within them, and the spirit of spontaneous
breaking is completely banished out of the fundamental theory~\footnote{
Nevertheless the idea of spontaneously broken symmetry is widely used when
constructing low-energy phenomenological models. We break in them, however,
not the symmetries of fundamental interactions, but effective gauge
symmetries, which are classificational at the level of fundamental theory.}.
This way leads us to building fundamental gauge group as a product
of a few simple groups, gradually (with energy decreasing) transiting into
confinement state~\footnote{
Within the built scheme technicolor concept are thought to be intermediate
relatively to GUTs and preon models: massive gauge bosons in it are
considered as fundamental, but Higgs mechanism isn't accepted, and
confined technicolor interaction provides techniquarks coupling into particles
which spontaneously break the gauge symmetries
}.

In recent years a few quark inner structure evidences were obtained
(\cite{1}-\cite{3}). That brings us to the necessity of extremely
sharp determination of low-energy preon structure predictions.
Among them is existence of leptoquark resonance, possible discovery of which
in DESY in the beginning of 1997 year caused a new wave of preon models
researches.

Existence of low-energy limit of SM bounds the number of different
preons within the model, which then helps to build a minimal
SM extention providing a possibility of phenomenological
preon structure effects calculation.
In that way, tiny ratio of weak interaction intensities for right- and
left-chiral fermions allows us to draw a proper conclusion on a great
difference of confinement scales for left- and right-chiral quarks and leptons.
This makes reasonable for us to consider right-chiral quarks and leptons structureless.

We follow the standard for preon models concept of weak interactions
interpreted as an exchange with universal spinor preons, which
are contained in all left-chiral fermions (we'd like to note
a great similarity to meson couplings in effective low-energy
barion-meson theories). This approach unavoidably introduces an extra
massive vector boson (so called $Z'$-boson). Its current mass constraint
is $m_{Z'}>340GeV$.

Within a built in such a way model quarks and leptons are
composed of scalar and spinor preons. The later are universal
preons, described in the previous paragraph and providing weak interaction.
Existence of fundamental scalar particles, however, introduces triple and quadruple
vertices, corresponding to new fundamental interactions (the vertices
are not forbidden by either renormalizability or gauge symmetry).
These strange interactions can be reduced
to the gauge ones by declaring supersymmetry or prohibited after adding
one more matter structural level -- subpreon.
In this paper we use the second of described possibilities: each
scalar preon is composed of two spinor subpreons.
This way leads us to a chiral model: in fundamental theory we have only
massless chiral spinor fields (and the gauge ones) --
other spin particles production and mass-gaining become possible due to
non-pertubative, QCD-like vacuum condensates.

\begin{figure}[t]
\begin{picture}(0,0)
\put(10,-51){\epsfbox{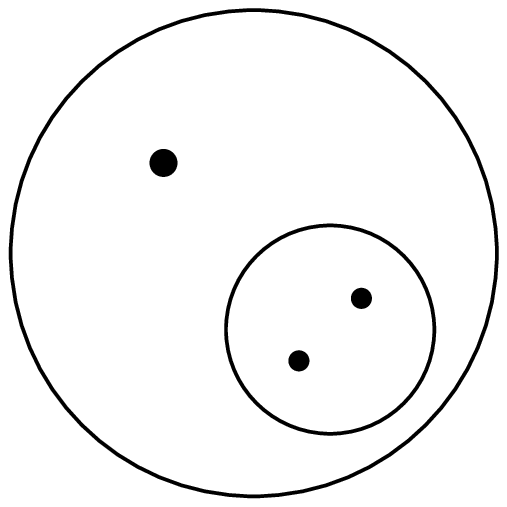}}
\put(90,-51){\epsfbox{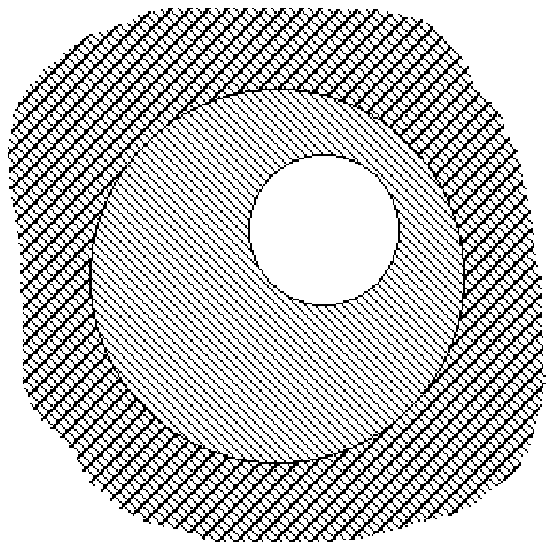}}
\put(38,-58){a}
\put(119,-58){b}
\end{picture}
\vspace{6cm}
\caption{a) a draft image of quark excitation within the examined model; b)
vacuum condensates distributions within a quark excitation.}
\label{quark}
\end{figure}

Analogously to hadron physics we suppose preons are confined in quarks,
and subpreons -- in preons. The confinement is provided by metacolor
(gauge group $SU_{mc}(N_{mc})$) and submetacolor (group
$SU_{smc}(N_{smc})$) interactions respectively. The suggestion is perhaps
too strong, but it's quite reasonable to believe that low-energy predictions
aren't sensitive to the exact mechanism of keeping preons inside quarks
and (still more evident) subpreons in preons. The resulting image of
left-chiral quark (or lepton) is shown at Fig.\ref{quark}.

In these paper our main hypothesis while building the model is
the forms of vacuum averages of fundamental fields. When they are
fixed our phenomenological notions and analogies to hadron physics
completely determine excitations spectrum.

In accordance with hadron level, there are two main types of vacuum
excitations in preon models:

1) so-called ``bags'' -- proton-like regions of molten preon-metagluon
condensate, made stable by valent preons.

2) collective excitations -- pion-like waves, generated with breaking
of fundamental fields correlations, which are inherent in vacuum state.

There is an important inner consistency criterion in a theory established in
the aforesaid way. In it all vector particles are build of chiral
preons and, consequently, initially massless. The only way for them
to gain a non-zero mass is non-pertubative vacuum interactions.
Massless particles (contrarily to massive ones) have only two
independent polarization components. That is why, condensate interactions
must also provide a transformation of each vector excitation appeared within
the theory into a scalar one, which has the same quantum numbers.
It is perceived as the third
necessary polarization component. Thus, within the build spectrum,
for each vector excitation we need to find a scalar one with the
same quantum numbers and build diagrams of transformation processes
through vacuum interactions. We can use some sort of condensate forms
only if within the spectrum corresponding to them the established condition
is satisfied (we also must require simplicity and
lack of contradictions with present experimental data).

After the spectrum of excitations is analyzed, we can extract the minimal
set of fields which are to be included into the low-energy model. As it has
been shown, the set can be ``put'' into the frameworks of
an extended SM with the group $U(1)\times U(1)\times SU(2)$.
The symmetry breaking in it is provided with effective Higgs mechanism,
which {\sl is not} considered in our model as fundamental.
It is carried out of preon model, that the build gauge theory must
be extended with some specific interactions of quarks and leptons with scalar
leptoquark, which are the most easily observable low-energy predictions of
preon structure existence. (When interpreted as a leptoquark effect
DESY data give it's mass of about $200GeV$). We can, after that, calculate
the influence of leptoquark resonance on the cross-section of
proton-positron scattering within the extended model.

In this paper we try to build a consistent model which contains all the
above-introduced concepts.
In section 1 we construct condensate forms and excitations spectra of
subpreon and preon levels and use phenomenological method of preon
diagrams to prove inner consistency of the build scheme.
Section 2 is dedicated to building effective low-energy
gauge theory with non-renormalizable leptoquark articles.
In section 3 we calculate cross-section of elementary $e^+d$-scattering
for processes with neutral and charged current and analyze some empirical
features of the obtained results.

\section{Preon phenomenology}

\newcommand{\scalpre}{\stackrel{\scriptscriptstyle +}{\varphi}\rule{-3pt}{0mm}{}}

\subsection{Preon types and gauge structure of the theory}

When constructing the set of elementary objects of the theory we aim
to satisfy two usual preon ideology postulates:

1) All fundamental interactions correspond to exact gauge symmetries.

2) Elementary fields are representations of the gauge groups.

In accordance with the first thesis, described in Introduction preon structure
and the necessity of QED and QCD groups presence we build the fundamental
gauge group in the form:

\[G=U(1)\times SU_c(3)\times SU_{mc}(N_{mc})\times SU_{smc}(N_{smc})\]
When symmetries $S_{smc}(N_{smc})$ and, then, $SU_{mc}(N_{mc})$ successively
transit into hidden state we are drawn into a usual quark-lepton level.
Fields, corresponding to the four factors of the gauge group $G$
are hereafter designated as $S_\mu , G_\mu^n, B_\mu^\omega, C_\mu^\Omega$;
their strength tensors are $S_{\mu\nu}, G_{\mu\nu}^n, B_{\mu\nu}^\omega,
C_{\mu\nu}^\Omega$.

There are three subpreon fields within the theory: universal right-chiral
subpreon $x_R^{\rho\alpha}$, which is included into all scalar preons,
and left-chiral subpreons  $Q_{La}^{\rho i}$ and $L_{Ll}^\rho$,
necessary for building quark and lepton preons respectively.
(Here index $\rho$ is associated with submetacolor interaction , $\alpha$ --
with metacolor, $i$ -- with chromodinamic, $a$ and $l$ -- flavour indices
of quarks and leptons; all the listed fields are transformed after
fundamental (when having two indices -- after bifundamental) representations
of corresponding groups).
If we move through the build hierarchy backwards from quark-lepton level
and require anomalies cancellation at each step, electric charges of the
fields $x$, $Q$ and $L$ are absolutely determined and equal to 0, 1/6 and -1/2.

Using the introduced notation we can write fundamental Lagrangian as follows:

\[L=-\frac14\sum_G{\bf G}_{\mu\nu}{\bf G}^{\mu\nu}
    +\sum_f i\bar f\hat Df\]
where $\sum_G$ means summation over gauge groups, and $\sum_f$
-- over all right- and left-chiral fermion fields.

Note, that unless postulates 1) and 2) are violated Lagrangian is automatically
invariant under transformation of an enormously rich global flavour group,
including independent flavour mixing of
right- and left-chiral leptons, left-chiral quarks and up and down
right-chiral quarks separately. We probably can reduce the flavour group,
by introducing some correlations of right- and left-chiral quarks and leptons
at preon level. Nevertheless, even when it's done the group is surely still
monstrously large.

For the reason that all fundamental fields are massless and all fundamental
interactions are gauge ones with exact symmetries the above-written Lagrangian
doesn't contain any dimensional parameters. They, however, must appear in
quantum theory because of non-pertubative vacuum condensates occurence
(which is necessary in a non-abelian theory)~\footnote{
When energies are far above the confinement scales and pertubative theory is
applicable, dimensional parameters appear after renormalization due to
dimensional transmutation effect. As it is shown by QCD exploring experience,
these parameters are equal to the characteristic scales of non-pertubative
vacuum by the order of magnitude.
}.
That is exactly the way, in which the fundamental scale hierarchy

\[\Lambda_1\gg\Lambda_{smc}\gg\Lambda_{mc}\gg\Lambda_{c}\]
appears. Here $\Lambda_{c}=100MeV$ -- usual QCD dimensional parameter.
The written scale takes a determining part in the formation of excitations
spectrum and properties. Quantities $\Lambda_{c}$, $\Lambda_{mc}$ and
$\Lambda_{smc}$ charecterize deconfinement energies of corresponding interactions.
For QED dimensional parameter, $\Lambda_1\sim 10^{19}GeV$, it only can
be said that it's, probably, the boundary of our most general notions of matter
structure applicability.

Quark and lepton scalar preons,
$$
\scalpre_a^{\alpha i}=\bar x_R^{\rho\alpha} Q_{La}^{\rho i} \quad \mbox{and} \quad
\scalpre_l^\alpha=\bar x_R^{\rho\alpha} L_{Ll}^{\rho},
$$
are constituted of spinor subpreons and have charges of 1/6 and -1/2.
We also need some universal left-chiral preons at the level of submetacolor
confinement: it is exchange with them which provides usual weak interaction.
It's not enough to introduce a single universal preon, for in the case
we'd only be able to build one vector boson. The minimal number
of weak interaction providing fields in physically consistent theory is two:
preons $U_L^{\alpha}$ and $D_L^{\alpha}$, holding the charges of 1/2 and -1/2.
``Chemical formulas'' of quark and lepton fields are:

\[u_{La}^i=U_L^\alpha\scalpre_a^{\alpha i} \qquad
  d_{La}^i=D_L^\alpha\scalpre_a^{\alpha i}\]
\[\nu_{Ll}=U_L^\alpha\scalpre_l^\alpha \qquad
  e_{Ll}=D_L^\alpha\scalpre_l^{\alpha}\]

The four vector bosons contained in the theory are composed of universal
preons.

\[
\begin{array}{l}
W_{\mu}^+=\bar D_L^{\alpha} \gamma_{\mu} U_L^{\alpha} \qquad
W_{\mu}^-=\bar U_L^{\alpha} \gamma_{\mu} D_L^{\alpha}
\\
\rule{0pt}{8mm}
W_{\mu}^3=\displaystyle\frac{1}{\sqrt 2}(\bar U_L^{\alpha} \gamma_{\mu} U_L^{\alpha}-
                             \bar D_L^{\alpha} \gamma_{\mu} D_L^{\alpha})
\\
\rule{0pt}{8mm}
W_{\mu}^0=\displaystyle\frac{1}{\sqrt 2}(\bar U_L^{\alpha} \gamma_{\mu} U_L^{\alpha}+
                             \bar D_L^{\alpha} \gamma_{\mu} D_L^{\alpha})
\end{array}
\]

Existence of an extra effective-gauge boson seems to be
an important prediction of the developed theory.
Its empirical consequences can be examined within the formalism which is
to be build in section 2.

\subsection{Submetacolor confinement level}

\setcounter{equation}{0}
\renewcommand{\theequation}{1.2.\arabic{equation}}

As it has been already said, submetacolor confinement is provided by
$SU_{smc}(N_{smc})$ gauge interaction. Condensate forms must be assumed
as a hypothesis (criteria of its verification were described in
Introduction). When making the choice we are guided by the following
heuristic recipe: first we build all sorts of Lorenz-invariant forms,
then construct of them non-abelian gauge groups singlets; the
singlets are used for condensate expressions formation, which are
invariant under $U(1)$-transformations. The performed sequence is
produced by our general notion of fundamental symmetries hierarchy.
More than that, requirement of throwing off all Lorenz non-invariant forms
at the first stage of vacuum condensate construction follows from our wish
not to have any low-energy excitations with high spin.
After the assumption of the described principle we come to unique set of
vacuum forms:

\begin{equation}
\begin{array}{l}
\langle 0|C_{\mu\nu}^\Omega C_\Omega^{\mu\nu}|0 \rangle =C_0
\\
\rule{0pt}{5mm}
\langle 0|(\bar x_R^{\rho \alpha} Q_{La}^{\sigma i})
          (\bar Q_{Lb}^{\sigma i} x_R^{\rho \alpha})|0 \rangle =C_{ab}^{(Q)}
\\
\rule{0pt}{6mm}
\langle 0|(\bar x_R^{\rho \alpha} L_{Ll}^\sigma)
          (\bar L_{Lm}^\sigma x_R^{\rho \alpha})|0 \rangle =C_{lm}^{(L)}
\end{array}
\label{sub_preon_cond}
\end{equation}
where $C$ matrices can be chosen diagonal.

Analogously to QCD, where the characteristic scale of quark condensates is
about one and a half times less than gluon ones, the ratios
$$
\Upsilon^{(Q)}=\frac{\sqrt[6]{\left\| C_{ab}^{(Q)}\right\|}}
                    {\sqrt[4]{C_0\vphantom{\left\| C\right\|}}}
\qquad \mbox{and} \qquad
\Upsilon^{(L)}=\frac{\sqrt[6]{\left\| C_{lm}^{(L)}\right\|}}
                    {\sqrt[4]{C_0\vphantom{\left\| C\right\|}}}
$$
in our theory must be less then one. The reason is that quark (and, as well,
preon) condensates are induced: quantum tunneling between
topologically different states of the field system produces gluon fields
fluctuations, which at their turn excite right- and left-chiral fields.
Forasmuch as QCD is a chiral theory, induced quark fluctuations distributions
are the same. A finite width of the distributions, however, makes
characteristic scale of quark condensate (which appears after averaging
of right- and left-chiral fluctuations product) is less than
gluon condensate scale factor.

Applying of the same approach to condensate forms (\ref{sub_preon_cond})
leads to even greater vacuum energies split.
Subpreon condensates are of fourth degree by fields and the averaging
must get a less result than in QCD case.
Thus, built in the previous paragraph dimensionless ratios with a high
probability can come close to one order of magnitude.

It seems to be an extremely important question, that of flavour symmetry
breaking with condensate forms (\ref{sub_preon_cond}). Heuristic analysis
of the problem shows that the symmetry can only be reduced to $SU(2)$.
The described picture of condensate inducing is flavour invariant.
The suggestion of spontaneous breaking, however, brings to the necessity
of instability development in the inducing processes. It is this instability
which leads to vacuum averages matrix flavour symmetry breaking.
Analogously to condensed matter physics state of affair, in low-symmetry phase
we have a domain with explicit dynamic equality of flavours (which took place
before symmetry breaking), the domain in which all the elements of vacuum
averages matrices $C$ are equal. When diagonalize the matrices we obtain
$C\sim {\rm diag}(3,0,0)$, which corresponds to the domain in which
symmetry breaking character is explicit~\footnote{
This, to tell the truth, a bit unclear statement can be confirmed by the
following example taken from ferromagnetism physics. In ferromagnetic
low-energy phase there is a domain with explicit equality of $x$, $y$ and
$z$-directions ($M\sim(1,1,1)$), and the domain in which symmetry breaking
down to $O(2)$ is evident($M\sim(1,0,0)$).}. We can also add some
isotropic (i.e. proportional to the identity matrix) part to the built vacuum
averages. It's not too difficult to see, that the obtained vacuum shift
matrices are $SU(2)$-invariant.

The above stated group theory arguments and a certain fact that first generation
is almost massless, brings us to the following conclusion on diagonal
elements of $C$-matrices values:

$$C_{11}^{(Q)}\ll C_{22}^{(Q)}=C_{33}^{(Q)}\quad and \quad
C_{11}^{(L)}\ll C_{22}^{(L)}=C_{33}^{(L)}.$$
This is quite a provocative result because small values of $C_{11}^{(Q)}$ and
$C_{11}^{(L)}$) can, in principle, lead to the appearance of small mass
excitations, different from scalar preons. It will be shown, that this
an unpleasant possibility isn't realized within the theory due to
submetagluon condensate presence.

We use the following phenomenological notions when obtaining the excitations
spectrum: a bag can be composed of any few (for low-energy excitations)
particles forming a colorless by submetacolor object with some definite
Lorenz-transformation properties. Such a system of particles stabilizes
condensate cavity. Besides that, we can destroy some fields correlation
intrinsic to (\ref {sub_preon_cond}) condensate and formally expressed
in index contractions (providing vacuum forms to be singlets).
This can be performed by undoing some contractions or changing
(\ref {sub_preon_cond}) vacuum averages. Thus collective excitations come.
Applying the stated recipes to (\ref {sub_preon_cond}) gives the following
spectrum:

1. Vector bags:

\begin{equation}
\begin{array}{l}
V_\mu^{\alpha \beta}=(\bar x_R^{\rho\alpha} \gamma_\mu x_R^{\rho\beta})=
                     (X_\mu , V_\mu^\omega)
\\
\rule{0pt}{5mm}
V_{\mu \rule{6pt}{0mm} ab}^{\rule{2pt}{0mm} ik}=
                  (\bar Q_{La}^{\rho i} \gamma _\mu Q_{Lb}^{\rho k})=
                     (V_{\mu \rule{3pt}{0mm} ab} ^{(Q)},
                      V_{\mu \rule{3pt}{0mm} ab}^{\rule{2pt}{0mm} n})
\\
\rule{0pt}{5mm}
V_{\mu \rule{3pt}{0mm} lm}^{(L)}=
                (\bar L_{Ll}^\rho \gamma_\mu L_{Lm}^\rho)
\\
\rule{0pt}{5mm}
\chi_{\mu \rule{2pt}{0mm} al}^{\rule{4pt}{0mm} i}=
     (\bar Q_{La}^{\rho i} \gamma _\mu L_{Ll}^\rho)
\end{array}
\label{sub_preon_vector_bags}
\end{equation}
in parentheses there are enumerated irreducible representations
of survived gauge symmetries contained within each excitation field.
The last field is subpreon vector leptoquark.

2. Scalar collective excitations (other than scalar preons):

\begin{equation}
\begin{array}{l}
\Psi^{\alpha \beta \rule{7pt}{0mm} ik}_{\rule{9pt}{0mm} ab}=
                              (\bar x_R^{\rho \alpha} Q_{La}^{\sigma i})
                              (\bar Q_{Lb}^{\sigma k} x_R^{\rho \beta})=
 (\Psi_{ab}, \Psi_{ab}^\omega, \Psi_{ab}^n,
  \Psi^{\omega \rule{7pt}{0mm} n}_{\rule{4pt}{0mm} ab})
\\
\rule{0pt}{6mm}
\Phi^{\alpha \beta}_{lm}=(\bar x_R^{\rho \alpha} L_{Ll}^\sigma)
                         (\bar L_{Lm}^\sigma x_R^{\rho \beta})=
 (\Phi_{lm}, \Phi_{lm}^\omega)
\\
\rule{0pt}{6mm}
\chi_{\rule{3pt}{0pt} al}^{i \rule{6pt}{0mm} \alpha\beta}=
                            (\bar x_R^{\rho \alpha} Q_{La}^{\sigma i})
                            (\bar L_{Ll}^\sigma x_R^{\rho \beta})=
 (\chi_{al}^i, \chi_{al}^{i \omega})
\\
\rule{0pt}{6mm}
C=C_{\mu \nu}^\Omega C_\Omega^{\mu \nu}
\\
\rule{0pt}{5mm}
a_C=e^{\mu \nu \rho \sigma}C_{\mu \nu}^\Omega C_{\rho \sigma}^\Omega
\end{array}
\label{sub_preon_scalar_exc}
\end{equation}

The last three particles are scalar subpreon leptoquark, submetaglueball
and pseudosubmetaglueball (submetaaxion). All the built excitations can be
expanded into representations of the reminder flavour group
$SU_{fam}^{(Q)}(2)\times SU_{fam}^{(L)}(2)$. When it's done, fields with
$ab$ and $lm$ indices are split into two singlets, two fundamental and a
adjoined representation of a corresponding $SU_{fam}(2)$ group.
As for leptoquark fields, their flavour group irreducible parts are
a both flavour group singlet, $SU_{fam}^{(Q)}(2)$ fundamental representation,
$SU_{fam}^{(L)}(2)$ fundamental and a bifundamental representation of
the whole reminder flavour group.

\begin{figure}[t]
\begin{picture}(0,0)
\put(20,-32){\epsfbox{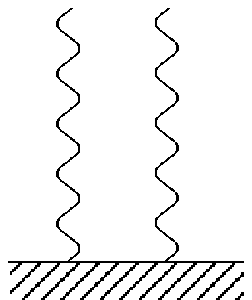}}
\put(65,-32){\epsfbox{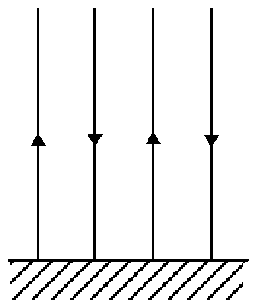}}
\put(112,-32){\epsfbox{pre_cond.eps}}
\put(26.5,0){$C_{\mu\nu}^\Omega$}
\put(36,0){$C^{\mu\nu}_\Omega$}
\put(69,0){$\bar x_R^{\rho\alpha}\rule{-1pt}{0mm}
            Q_{La}^{\sigma i}\rule{-2pt}{0mm}
            \bar Q_{Lb}^{\sigma i}
            x_R^{\rho\alpha}$}
\put(116,0){$\bar x_R^{\rho\alpha}\rule{-1pt}{0mm}
             L_{Ll}^\sigma\rule{-1pt}{0mm}
             \bar L_{Lm}^\sigma\rule{-1pt}{0mm}
             x_R^{\rho\alpha}$}
\end{picture}
\vspace{100pt}
\caption{Effective vertices, corresponding to interactions with subpreon level
vacuum condensates.}
\label{vac_vert}
\end{figure}

It's of a considerable importance to know (\ref {sub_preon_vector_bags})
and (\ref {sub_preon_scalar_exc}) abilities for mutual transformations and
interactions with (\ref {sub_preon_cond}) condensate. Only by carrying out the
answer, we can examine which combinations of (\ref {sub_preon_scalar_exc}) fields
are perceived as third independent polarization components.
It's impossible to achieve a quantitative answer to the question within
modern field theory. We, however, can reveal some important processes
features through phenomenological technique of preon diagrams, which
are graphic images of processes with (\ref {sub_preon_vector_bags}) and
(\ref {sub_preon_scalar_exc}) particles. The later are visualized as
composed of preons interacting in accordance with fundamental
$U(1)\times SU_c(3)\times SU_{mc}(N_{mc})\times SU_{smc}(N_{smc})$-Lagrangian.
Within such an approach excitation-condensate interactions are described
by special vertices, drawn at Fig.~\ref{vac_vert}.
In hadron physics a similar technique were established in a classical paper
\cite{vacuum}.

We examine only confinement-irreducible diagrams, i.e. those
which can't be decomposed into parts, external lines of which form
colorless (by submetacolor) groups. (That sort of diagrams corresponds
to elementary low-energy processes). More than that, it's quite evident
that external lines of each considered diagram must themselves form
colorless groups (due to confinement). Size of the necessarily examined
diagrams set is strongly reduced by the two stated requirements, and so
we can do a complete enumeration. We don't draw all the essential diagrams
here because of quite a big number of them. It's only possible to perform
some important diagram examples and the results obtained for
mass-gaining, mixing and third polarization components production processes.

Due to both fundamental and low-energy Lagrangians are gauge-invariant,
only fields transforming after the same representation are able for
mutual conversion. This divides (\ref {sub_preon_vector_bags}) and
(\ref {sub_preon_scalar_exc}) into subsets of particles, which mix and
form polarization components only within the group:
\vspace{-1mm}
\begin{tabbing}
\quad\=III.\quad\=$(X_\mu, V_{\mu \rule{3pt}{0mm} aa}^{(Q)},
                    V_{\mu \rule{3pt}{0mm} ll}^{(L)}) \qquad$\=
        $(C, a_C, \Psi_{aa}, \Phi_{ll})$\rule{10pt}{0mm}-\rule{10pt}{0mm}
        \=comment\kill
\>I.\>$(X_\mu, V_{\mu \rule{3pt}{0mm} aa}^{(Q)},
        V_{\mu \rule{3pt}{0mm} ll}^{(L)})$\>
       $(C, a_C, \Psi_{aa}, \Phi_{ll})$
    \rule{10pt}{0mm}-\rule{10pt}{0mm}\>(no summation over $a$ and $l$ !)\\
\rule{0pt}{5mm}
\>II.\>$V_{\mu \rule{3pt}{0mm} ab}^{(Q)}$\>$\Psi_{ab}$\>$(a \ne b)$\\
\rule{0pt}{5mm}
\>III.\>$V_{\mu \rule{3pt}{0mm} lm}^{(L)}$\>$\Phi_{lm}$\>$(l \ne m)$\\
\rule{0pt}{5mm}
\>IV.\>$V_\mu^\omega$\>$(\Psi_{aa}^\omega, \Phi_{ll}^\omega)$\\
\rule{0pt}{5mm}
\>V.\>$V_{\mu \rule{4pt}{0mm} ab}^{\rule{2pt}{0mm} n}$\>$\Psi^n_{ab}$\\
\rule{0pt}{5mm}
\>VI.\>$\chi_{\mu\rule{2pt}{0mm} al}^{\rule{4pt}{0mm} i}$\>
       $\chi^i_{al}$
\end{tabbing}
\vspace{-4pt}

\begin{figure}[t]
\begin{picture}(0,0)
\put(15,-34){\epsfbox{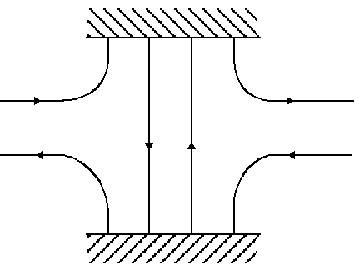}}
\put(63,-40){\epsfbox{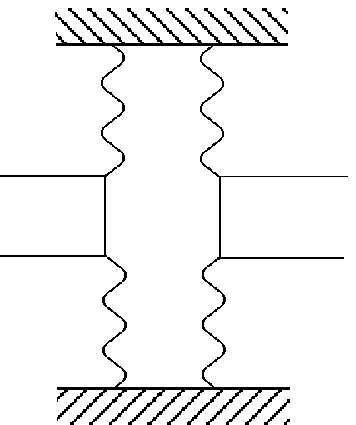}}
\put(112,-35){\epsfbox{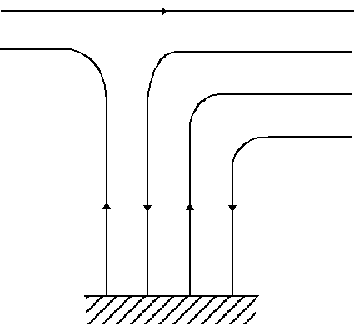}}
\put(32,-48){a}
\put(79,-48){b}
\put(128,-48){c}
\end{picture}
\vspace{150pt}
\caption{``Skeletons'' of some essential preon diagrams, which are used when
analyzing excitations features: vector boson mass-gaining -- a) due to
interaction with preon condensate, b) due to interaction with gluon
condensate; c) third independent polarization component production.
Diagram b) contains three confinement-irreducible parts:
vector particle transformation into submetaglueball (metaglueball),
mass-gaining by the later and its reverse transformation into
a vector particle.}
\label{processes}
\end{figure}

Below there are main results of preon diagrams analysis for each of the
groups (the most essential mass-gaining and third components production
diagrams are performed at Fig.~\ref{processes})

I. Preon diagrams provide all third components production.
   All vector bosons are massive (for first flavour the main contribution
   into masses is given by diagram of interaction with submetagluon
   condensate at Fig.~\ref{processes}b). As it has been mentioned,
   lack of massless colorless object within the theory is of
   a great importance for not having contradiction with present
   experimental data.

II. Mass-gaining mechanism works. There are some less massive particles,
    but their masses aren't catastrophically small
    (of order $\sqrt{C_{11}/C_{22}}$ relatively to submetacolor confinement
    scale).

III. Everything is analogous to II.

IV. Third components producing is provided. All vector particles are massive.

V. Vacuum interactions provide third components gaining.
There are some almost massless particles among vector fields of the group
(which together form massive gluons octet).
Considerable mass split of heavy gluons must carry to chromodinamic interactions
flavour symmetry breaking at high energies.

VI. Likely to the previous case, some of yet unobservable colored objects
are almost massless. Third polarization components are bounded
to vector particles with condensate interactions.

Thus, Introduction consistency criterion is satisfied,
and the choice of (\ref {sub_preon_cond}) confirmed in the sense of
theory inner consistency. There are also no rough  and evident
experimental data contradictions (e.g. extra massless observable
objects presence, unnatural spin qualities fields {\sl etc.}).

It's quite reasonable to include all the constructed compound fields
into an effective low-energy theory of submetacolor confinement level.
The task is not too difficult, because all
(\ref{sub_preon_vector_bags}) and (\ref {sub_preon_scalar_exc}) fields
decay into scalar preons and the later have small masses.
The last statement isn't of a big evidence:
there are two main arguments for it -- the first is due to hypothesis
of chiral character of the model, the last -- due to molten condensate
regions localization character.

Our first argument arises from a strong notion of the fact that the resulting
compound field $q_{La}^i$ must be massless. Because of contained in it
scalar preons have themselves a non-zero mass, we need some subtle dynamic
adjustment, which provides vanishing of total rest energy of quark excitation.
This happens due to summary scalar preon energy and condensate melting energy
compensation with negative energy of preons' interaction. Such an adjustment
seems to be extremely improbable when scalar preon mass is big.

There are also some more plain arguments for small scalar preon mass:
there is a picture of vacuum condensates state in quark (lepton) excitation
at Fig.~\ref{quark}b (thick dashing corresponds to metacolor confinement
condensates, thin -- to submetacolor confinement ones).
We represent quark as a bag excitation, but within the present notions
any excitation is associated with a localized in some way region of molten
condensate. Consequently, our arguments are applicable independently upon
quark structure details. Dashing lack in inner part of Fig.~\ref{quark}b
marks exactly the region of molten condensate.

It's easy to see, that no one of (\ref{sub_preon_cond}) condensates provide
scalar preon mass-gaining (contrarily to the other built subpreon scalar
excitations). Fig.~\ref{quark}b shows that metacolor confinement
condensates, the only ones which are able to give a non-zero mass
to scalar preon, are molten in it's vicinity and so can't
fully execute their functions. This makes us think that scalar
preon masses are probably far less than metacolor confinement scale
(and undoubtedly not similar to submetacolor confinement scale, which
can be deduced from na\"\i ve reasoning).

Thus, eliminating all massive unstable objects, we have only quark and
lepton scalar preons and gauge fields to be included into a consistent
low-energy theory. When adding universal spinor preons which arise at
the next level of confinement hierarchy and fix the
$U(1)\times SU_c(3)\times SU_{mc}(N_{mc})$ gauge group we come to
well-known boson-fermion preon model (\cite{preon1}, \cite{preon2}).
We'd like to note that in this paper four-particle interactions
are not considered to be fundamental: it's easy to build preon diagrams
expressing this exotic processes inner structure in terms of
$U(1)\times SU_c(3)\times SU_{mc}(N_{mc})\times SU_{smc}(N_{smc})$ theory,
which doesn't contain any unnatural vertices.

\subsection{Metacolor confinement level}

\setcounter{equation}{0}
\renewcommand{\theequation}{1.3.\arabic{equation}}

Acting similarly to the previous section, we introduce the following
condensates of the gauge field and scalar and spinor preon fields to
provide confinement~\footnote{
We don't take into account condensates of any fields but those of boson-fermion model
and those which considered as fundamental within the paper. Condensates
of other excitations built in previous section are negligibly small because
of their huge masses.}:

\begin{equation}
\begin{array}{l}
\langle 0|B_{\mu \nu}^\omega B_\omega ^{\mu \nu}|0 \rangle = C_B
\\
\rule{0pt}{5mm}
\langle 0|\scalpre_a^{\alpha i}\varphi_b^{\alpha i}|0 \rangle =C_{ab}^{\varphi _{(Q)}} \rule{61pt}{0mm}
\langle 0|\scalpre_l^\alpha\varphi_m^\alpha |0 \rangle =C_{lm}^{\varphi _{(L)}}
\\
\rule{0pt}{6mm}
\langle 0|(\bar U_L^\alpha x_R^{\rho \beta})
          (\bar x_R^{\rho \beta} U_L^\alpha)|0 \rangle =C_U^{(1)} \qquad
\langle 0|(\bar U_L^\alpha x_R^{\rho \alpha})
          (\bar x_R^{\rho \beta} U_L^\beta)|0 \rangle =C_U^{(2)}
\\
\rule{0pt}{6mm}
\langle 0|(\bar D_L^\alpha x_R^{\rho \beta})
          (\bar x_R^{\rho \beta} D_L^\alpha)|0 \rangle =C_D^{(1)} \qquad
\langle 0|(\bar D_L^\alpha x_R^{\rho \alpha})
          (\bar x_R^{\rho \beta} D_L^\beta)|0 \rangle =C_D^{(2)}
\end{array}
\label{preon_cond}
\end{equation}

The fact, that we successively excluded all the submetacolor confinement
level excitation with only exception for boson-fermion model fields,
can make subpreon level to seem unnecessary and unnatural.
Such arguments, however, doesn't hurt the built hierarchy scheme,
for $C_U$ and $C_D$ cannot be composed from boson-fermion model fields only.
That is one more (perhaps, even more strong) theoretical approval of
subpreon level necessity -- excitations spectrum, which both passes through
our consistency criterion and doesn't contradict to known experimental facts,
can't be built without its assuming.

As for (\ref {preon_cond}) condensates characteristic scales ratio, everything
said in the previous section is applicable. More than that, $U$- and
$D$-condensates non-zero values are caused by correlations of
induced fluctuations of different scales, and, consequently, their energy
split has to be even greater than in case of subpreon level.
In low-energy theory $U$- and $D$-condensates correspond to vacuum averages
of Higgs fields and so they are about the scale of effective gauge
$SU(2)$-symmetry breaking, which is experimentally known
($\Lambda_{SM}\approx 100GeV$). When the arguments are revealed,
small value of this quantity relatively to the presupposed metacolor
deconfinement scale seems to be a natural theory's consequence, but
not a destroying its orderliness paradox.

We construct excitations spectrum using principles established in previous
section:

1. Vector bags:

\begin{equation}
\begin{array}{l}
W_\mu^+=(\bar D_L^{\alpha}\gamma_\mu U_L^{\alpha})
\\
\rule{0pt}{5mm}
W_\mu^{(U)}=(\bar U_L^{\alpha}\gamma_\mu U_L^{\alpha})
\\
\rule{0pt}{5mm}
W_\mu^{(D)}=(\bar D_L^{\alpha}\gamma_\mu D_L^{\alpha})
\end{array}
\label{preon_vector_bags}
\end{equation}

2. Spinor bags -- quark and leptons.

3. Scalar collective excitations:

a) excitations of scalar preon condensates:

\begin{equation}
\begin{array}{l}
\Psi_{ab}^{ik}=(\scalpre_a^{\alpha i}\varphi_b^{\beta k})=(\Psi_{ab}, \Psi^n_{ab})
\\
\rule{0pt}{5mm}
\Phi_{lm}=(\scalpre_l^\alpha\varphi_m^\alpha)
\\
\rule{0pt}{5mm}
\chi^i_{al}=(\scalpre_a^{\alpha i}\varphi_l^\alpha)
\end{array}
\label{preon_scalar_exc}
\end{equation}
the later excitation is scalar preon leptoquark, observed as a resonance
in proton-positron scattering.

b) excitations of spinor preon condensate

\begin{equation}
\begin{array}{c}
W^+=(\bar D_L^\alpha x_R^{\rho \beta})
    (\bar x_R^{\rho \beta} U_L^\alpha) \qquad
H^+=(\bar D_L^\alpha x_R^{\rho \alpha})
    (\bar x_R^{\rho \beta} U_L^\beta)
\\
\rule{0pt}{6mm}
W^{(U)}=(\bar U_L^\alpha x_R^{\rho \beta})
        (\bar x_R^{\rho \beta} U_L^\alpha) \qquad
H^{(U)}=(\bar U_L^\alpha x_R^{\rho \alpha})
        (\bar x_R^{\rho \beta} U_L^\beta)
\\
\rule{0pt}{6mm}
W^{(D)}=(\bar D_L^\alpha x_R^{\rho \beta})
        (\bar x_R^{\rho \beta} D_L^\alpha) \qquad
H^{(D)}=(\bar D_L^\alpha x_R^{\rho \alpha})
        (\bar x_R^{\rho \beta} D_L^\beta)
\end{array}
\label{univ_preon_exc}
\end{equation}
$W$ and $H$ particles form eight degrees of freedom,
which play the roles of four Goldstone and four Higgs bosons in effective
low-energy theory.

c) metaglueballs

\begin{equation}
B=B_{\mu\nu}^\omega B_\omega^{\mu\nu} \qquad
a_B=e^{\mu\nu\rho\sigma}B_{\mu\nu}^\omega B_{\rho\sigma}^\omega
\label{metaglueballs}
\end{equation}
add two more Higgs-Goldstone degrees of freedom, which allows to build
a theory with two doublets and a complex singlet.

Within the standard method, we divide excitations into groups capable for
mutual transformation:

I.    $W_\mu^+ \qquad W^+$

II. $(W_\mu^{(U)}, W_\mu^{(D)})\qquad (W^{(U)}, W^{(D)}, B, a_B)$
where we also need to take into account the ability of mixing II group
particles with I group bosons from the previous section.
Diagram analysis assures that third polarization components producing
mechanism works properly at this level too.

As it can be seen, the main degrees of freedom of metacolor
confinement level theory, i.e. quark-lepton model, can be caught
into a gauge theory with $U(1)\times U_W(1)\times SU_L(2)$-group,
two Higgs-doublets and a singlet transforming after $U_W(1)$ only.
Preon model predicts singlet vacuum shift to be far greater than doublet ones.
At preon level singlet corresponds to a mixture of excitations, the main
component of which is metaglueball; but vacuum average of the later is
nothing else but $C_B$, which, as it has been said above, strongly exceeds
all the other condensates. Such a state of affair causes a small ratio of
$Z$- and $Z'$-bosons masses, which, thus, unavoidably follows from
preon level existence.

Without any violation of gauge symmetry we can also add into the built
low-energy theory a non-renormalizable vertex, coupling first flavour scalar
leptoquark to fermions and Higgs particles. Extended in this way theory can be
used for resonance $e^+d$- and $u\bar\nu_e$-scattering calculations.

\section{Effective low-energy gauge model of quark-lepton level}

\subsection{$U(1)\times U(1)\times SU(2)$ gauge theory}

\setcounter{equation}{0}
\renewcommand{\theequation}{2.1.\arabic{equation}}

It's quite reasonable to try building a gauge low-energy theory.
First, we know that it must at least include a gauge theory with
$U(1)\times SU(2)$ group (SM namely). Second, requirement of
renormalizability of some model sector (formed by the lightest particles) and
presence of massive vector bosons among the particles which are to be included
brings us to the necessity of gauge structure.

Four massive vector bosons extracted from the metacolor level excitation
spectrum can be driven out of a $U(1)\times U_W(1)\times SU_L(2)$-theory
after spontaneous symmetry breaking to $U(1)$. Fields $S_\mu, W_\mu^0$ and
$W_\mu^i$ correspond to three simple gauge groups.
Higgs sector contains two $SU_L(2)$-doublets with equal $U_W(1)$-charges
and $U(1)$-charges which ratio is equal to -1 (the doublets are designated as $h_1$ and $h_2$).
Complex singlet $\sigma$ is transformed after $U_W(1)$ only;
its charge can be chosen in some different ways. When usual discrete symmetries
of two-doublet standard model (\cite{VZS}) are accepted the Higgs potential
structure, necessary for lack of massless Goldstone bosons within the theory,
puts the singlet $U_W(1)$-charge twice greater than doublets' ones.

We chose vacuum shifts in the form:

\begin{equation}
h_1={0 \choose \rho_1} \qquad
h_2={\rho_2 \choose \delta} \qquad
\sigma=\frac{\sigma_0}2
\label{vac_shift}
\end{equation}
where $\delta=0$ in massless photon phase (which only will be analyzed below).
Singlet vacuum shift can be made real by redefining the $\sigma$ field, which
hereafter is supposed to be done. When these two restrictions are satisfied
any vacuum shifts set can be reduced to the form (\ref {vac_shift})
by gauge transformations.

Applying unitary gauge conditions we obtain the following expressions for
the Higgs fields:

\begin{equation}
\begin{array}{c}
h_1=\left(
     \begin{array}{c}
     H^+\sin \varphi
     \\
     \rule{0pt}{4mm}
     \displaystyle
     \rho \sin\xi\cos\varphi+H_1^0+
     i\frac{\cos\xi\sin\varphi}
           {\sqrt{1-\sin^2\xi\cos^22\varphi}}A^0
     \end{array}
    \right)
\\
\rule{0pt}{15mm}
h_1=\left(
     \begin{array}{c}
     \displaystyle
     \rho \sin\xi\sin\varphi+H_2^0+
     i\frac{\cos\xi\cos\varphi}
           {\sqrt{1-\sin^2\xi\cos^22\varphi}}A^0
     \\
     \rule{0pt}{4mm}
     H^- \cos  \varphi
     \end{array}
    \right)
\\
\rule{0pt}{9mm}
\displaystyle\sigma=\frac{\rho\cos\xi}2+
       H_3^0+
       i\frac{\sin\xi\sin2\varphi}
             {\sqrt{1-\sin^2\xi\cos^22\varphi}}A^0
\end{array}
\label{unit_gauge}
\end{equation}
where $H^+, H_1^0, H_2^0, H_3^0$ and $A^0$ are charged and neutral Higgs
fields.
\vspace{-2mm}
$$\rho=\sqrt{\rho_1^2+\rho_2^2+\sigma_0^2},\quad
\xi=\arccos(\sigma_0/\rho),\quad \varphi=\arctan(\rho_2/\rho_1)$$.

\vspace{-7mm}
Supposing coupling constants equal to $g_1/2$ for $U(1)$, $g_W/2$ for $U_W(1)$
and $g_2$ for $SU_L(2)$ and substituting (\ref {unit_gauge}) expressions into
articles with covariant derivatives of Higgs fields, we obtain
$W_\mu^+$ mass and $A_\mu$, $Z_\mu$ and $Z'_\mu$ mass matrix in the form:

\begin{equation}
\begin{array}{c}
\displaystyle m_{W^+}=g_2\sqrt\frac{\rho_1^2+\rho_2^2}{2}
\\
\rule{0pt}{15mm}
\displaystyle (M^2)_{S, W^3, W^0}=\frac{\rho^2}2
\left(
\begin{array}{l}
\rule{15pt}{0mm} g_1^2\sin ^2\xi \rule{50pt}{0mm}
-g_1g_2\sin^2\xi \rule{25pt}{0mm}
-g_1g_W\sin^2\xi\cos2\varphi
\\
\rule{0pt}{8mm}
-g_1g_2\sin^2\xi \rule{60pt}{0mm}
g_2^2\sin^2\xi \rule{45pt}{0mm}
g_2g_W\sin^2\xi\cos2\varphi
\\
\rule{0pt}{8mm}
-g_1g_W\sin^2\xi\cos2\varphi \rule{15pt}{0mm}
g_2g_W\sin^2\xi\cos2\varphi \rule{50pt}{0mm}
g_W^2
\end{array}
\right)
\end{array}
\label{mass_matrix}
\end{equation}
When diagonalized, (\ref {mass_matrix}) gives massless electromagnetic field
and two massive vector boson fields:

\begin{equation}
\begin{array}{c}
\displaystyle m_Z^2=\frac{\rho^2}4
    \left(
    (g_1^2+g_2^2)\sin^2\xi+g_W^2-
    \sqrt{\left[
    (g_1^2+g_2^2)\sin^2\xi-g_W^2\right]^2-
    4(g_1^2+g_2^2)g_W^2\sin^4\xi\cos^22\varphi}
    \right)
\\
\rule{0pt}{8mm}
\displaystyle m_{Z'}^2=\frac{\rho^2}4
    \left(
    (g_1^2+g_2^2)\sin^2\xi+g_W^2+
    \sqrt{\left[
    (g_1^2+g_2^2)\sin^2\xi-g_W^2\right]^2-
    4(g_1^2+g_2^2)g_W^2\sin^4\xi\cos^22\varphi}
    \right)
\end{array}
\label{ZZ'mass}
\end{equation}
Because the quantity $m_Z/m_{Z'}$ is small we can expand this
expressions into series by $\sin^2\xi$:

\newcommand{\gsww}{\frac{g_1^2+g_2^2}{g_W^2}}
\begin{equation}
\begin{array}{l}
\displaystyle m_Z^2=\frac{m_{W^+}^2}{\cos^2\theta_W}\left(
                    1-\sin^2\xi\cos^22\varphi-
                    \gsww\sin^4\xi\cos^22\varphi+\cdots\right)
\\
\rule{0pt}{8mm}
\displaystyle m_{Z'}^2=\frac{g_W^2\sigma_0^2}{2\cos^2\xi}\left(
                       1+\gsww\sin^4\xi\cos^22\varphi+
                       \left(\gsww\right)^2\sin^6\xi\cos^22\varphi+\cdots\right)
\\
\rule{0pt}{8mm}
\displaystyle\frac{m_Z^2}{m_{Z'}^2}=\gsww\sin^2\xi
                    -\left(\gsww\right)^2\sin^4\xi\cos^22\varphi+\cdots
\end{array}
\label{ZZ'mass_approx}
\end{equation}
First article in $m_Z^2$ expansion gives a usual ratio of
$W^+$- and $Z$-bosons masses in SM. Due to equality of doublets
$U_W(1)$-charges (which follows from preon model) when doublets shifts are equal
($\cos2\varphi=0$) $W_\mu^0$ field doesn't mix with other fields
and SM mass ratio is sharply reproduced. Note, that $W^+$- and $Z$-bosons
masses, determined mainly by doublets shifts lie at Salam-Weinberg scale,
and $Z'$-boson mass, formed by singlet vacuum shift, must be of
about metacolor deconfinement energy scale.

We analyzed Higgs potential in the most general form allowed by
discrete and gauge symmetries:

\begin{equation}
\rule{73pt}{0mm}
\begin{array}{l}
\rule{-73pt}{0mm}
U(h_1,h_2,\sigma)=-\mu_1(h_1^+h_1)-\mu_2(h_2^+h_2)-\mu_3(\sigma^+\sigma)+
\\
\rule{0pt}{5mm}
+\lambda_{11}(h_1^+h_1)^2+\lambda_{22}(h_2^+h_2)^2+
\lambda_{33}(\sigma^+\sigma)^2+
\\
\rule{0pt}{5mm}
+\lambda_{12}(h_1^+h_1)(h_2^+h_2)
+\bar\lambda_{12}(h_1^+h_2)(h_2^+h_1)+
\\
\rule{0pt}{5mm}
+\tilde\lambda_{12}(h_1^+\tilde h_2)(\tilde h_2^+h_1)+
\\
\rule{0pt}{5mm}
+\lambda_{13}(h_1^+h_1)(\sigma^+\sigma)
+\lambda_{23}(h_2^+h_2)(\sigma^+\sigma)+
\\
\rule{0pt}{5mm}
+\nu\left[(h_1^+\tilde h_2)\sigma+(\tilde h_2^+h_1)\sigma^+\right]
\end{array}
\label{potential}
\end{equation}
where $\tilde h=i\tau_2h^+$.

Substitution (\ref{vac_shift}) into (\ref{potential}) and following
minimization brings us to vacuum averages equations:

\begin{equation}
\left\{
\begin{array}{l}
\displaystyle
-\mu_1\rho_1+2\lambda_{11}\rho_1^3+(\lambda_{12}+\tilde\lambda_{12})\rho_1\rho_2^2+
\frac14\lambda_{13}\rho_1\sigma_0^2-\frac12\nu\rho_2\sigma_0=0
\\
\rule{0pt}{8mm}
\displaystyle
-\mu_2\rho_2+2\lambda_{22}\rho_2^3+(\lambda_{12}+\tilde\lambda_{12})\rho_1^2\rho_2+
\frac14\lambda_{23}\rho_2\sigma_0^2-\frac12\nu\rho_1\sigma_0=0
\\
\rule{0pt}{8mm}
\displaystyle
-\mu_3\sigma_0+\frac12\lambda_{33}\sigma_0^2+
(\lambda_{13}\rho_1^2+\lambda_{23}\rho_2^2)\sigma_0+2\nu\rho_1\rho_2=0
\end{array}
\right.
\label{vac_equations}
\end{equation}

$H^+$ and $A^0$ masses and $H_1^0, H_2^0$ and $H_3^0$ mass matrix can be found
from (\ref{unit_gauge}) and (\ref{potential}):

\begin{equation}
\begin{array}{c}
\displaystyle m_{H^+}^2=(\bar\lambda_{12}-\tilde\lambda_{12})(\rho_1^2+\rho_2^2)+
\frac{\nu\sigma_0\cos^22\varphi}{\sin2\varphi}
\\
\rule{0pt}{8mm}
\displaystyle m_{A^0}^2=\nu\sigma_0\cos^2\xi\frac{\cos^22\varphi}{\sin2\varphi}
\\
\rule{0pt}{20mm}
(M^2)_{1,2,3}=\left(
\begin{array}{l}
\rule{10pt}{0mm}
\displaystyle4\lambda_{11}\rho_1^2+\frac12\nu\sigma_0\tan\varphi
\rule{30pt}{0mm}
\displaystyle2(\lambda_{12}+\tilde\lambda_{12})\rho_1\rho_2-\frac12\nu\sigma_0
\rule{15pt}{0mm}
\lambda_{13}\rho_1\sigma_0-\nu\rho_2
\\
\rule{0pt}{10mm}
\displaystyle2(\lambda_{12}+\tilde\lambda_{12})\rho_1\rho_2-\frac12\nu\sigma_0
\rule{20pt}{0mm}
\displaystyle4\lambda_{22}\rho_2^2+\frac12\nu\sigma_0\cot\varphi
\rule{30pt}{0mm}
\lambda_{23}\rho_2\sigma_0-\nu\rho_1
\\
\rule{0pt}{10mm}
\rule{20pt}{0mm}
\lambda_{13}\rho_1\sigma_0-\nu\rho_2
\rule{70pt}{0mm}
\lambda_{23}\rho_2\sigma_0-\nu\rho_1
\rule{35pt}{0mm}
\displaystyle\lambda_{33}\sigma_0^2+2\nu\frac{\rho_1\rho_2}{\sigma_0}
\end{array}
\right)
\end{array}
\label{higgs_mass}
\end{equation}

Pertubative theory applicability conditions helps to reveal some constraints
put on (\ref{potential}) parameters:
\vspace{-2mm}
\[\frac{\nu}{4\pi^2\sigma_0}\ll 1 \qquad \frac{\lambda}{4\pi^2}\ll 1.\]
Using these relations and $\sin\xi\ll 1$ inequality, we can approximately
(in first order by $\sin\xi$) diagonalize Higgs fields mass matrix.
Two of the obtained in this way masses, $\sqrt{\lambda_{33}}\sigma_0$ and
$\sqrt{\nu\sigma_0(\tan\varphi+\cot\varphi)/2}$, lie at the scale of metacolor
confinement, and the third (vanishing within the considered
approximation) -- at Salam-Weinberg scale.

To draw any conclusions on $H^+$ and $A^0$-bosons masses, we need to know
the quantity $\cos2\varphi$. It's quite obvious at preon level that it must
be small: $\cos2\varphi\sim\rho_1^2-\rho_2^2$ and doublet shifts are
determined by condensate forms $C_U$ and $C_D$ from (\ref{preon_cond}).
Taking into view that $C^{(1)}$ and $C^{(2)}$ expressions, corresponding
to $h_1$ and $h_2$ at low-energy level, have similar structure, we almost
unavoidably arrive to the conclusion that induced by similar processes
$C^{(1)}$ and $C^{(2)}$ condensates are equal to each other quite precisely.

We would like to stress, that sharp equality $\rho_1=\rho_2$ can't take place
within a consistent theory, because, in such a situation, $A^0$ boson turns out
to be massless. Thus, the quantity $\cos2\varphi$ which is determined
by doublets vacuum shift ratio is of a small non-zero value, and axial
neutral Higgs-boson mass is formed as a product of big singlet vacuum shift
$\sigma_0$ and small $\cos2\varphi$ and lie at a scale which cannot be
determined with a heuristic investigation.

Some remarks can also be made on the built theory predictions for the value of
$W^+$- and $Z$-bosons masses precision ratio, which experimental value is
according to SM predictions up to 1\% sharp. It can be seen from
(\ref{ZZ'mass_approx}) formula that relative deviation of $m_Z/m_{W^+}$ ratio
from SM predictions in our theory is proportional to $\sin^2\xi\cos^22\varphi$.
As it has been mentioned, $\cos2\varphi$ is a small quantity and
$\sin\xi\sim m_Z/m_{Z'}$. Thus, (due to carried out of preon model doublets
$U_W(1)$-charges equality) within the examined theory high sharpness of
SM predictions for precision ratio doesn't put any restrictions on $Z'$-boson
mass. Even when it's not too big, presence of small multiplier will make deviations
from SM predictions negligible.

\subsection{Low-energy theory expansion with non-renormalizable interactions}

\setcounter{equation}{0}
\renewcommand{\theequation}{2.2.\arabic{equation}}

In previous section we included into effective low-energy model the maximum
field set, which allows to build a renormalizable theory. However
DESY collider experiments make considerably important to establish a
calculation technique for scalar leptoquark engaging processes.
Preon diagrams analysis brings to a conclusion that all fermion-leptoquark
interactions are non-renormalizable (it's not a bit wondrous as
low-energy interactions are {\sl a priori} non-local).

It's easy to see that due to exactly interactions non-locality the number
of different low-energy vertices engaging leptoquark is infinite.
We'll try to point out their main features and consider some simplest
representatives.

Note, that within the excitations structure notions, developed in section
1, leptoquark interacts with left-chiral fermions only. It, however, can't
decay into left-chiral quark and left-chiral fermion due to momentum and
angular momentum conservation laws. We know two ways to overcome this obstacle.

\begin{figure}[t]
\begin{picture}(0,0)
\put(50,-42){\epsfbox{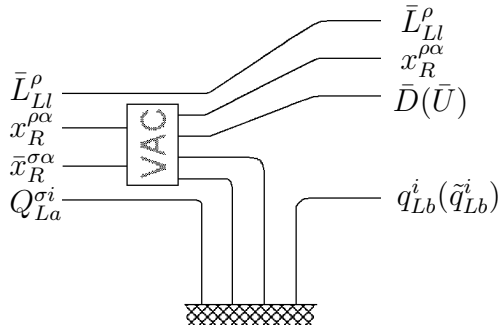}}
\put(43,-12){$\bar L_{Ll}^\rho$}
\put(43,-17){$x_R^{\rho\alpha}$}
\put(43,-22){$\bar x_R^{\sigma\alpha}$}
\put(43,-27){$Q_{La}^{\sigma i}$}
\put(95,-3){$\bar L_{Ll}^\rho$}
\put(95,-8){$x_R^{\rho\alpha}$}
\put(94,-13.5){$\bar D(\bar U)$}
\put(94.5,-26){$q_{Lb}^i(\tilde q_{Lb}^i)$}
\end{picture}
\vspace{122pt}
\caption{Effective sum of one class of leptoquark decay processes.}
\label{lptqdecay}
\end{figure}

1) The born left-chiral quark ``immediately'' interacts with the condensate,
providing quarks mass-gaining and transforms into a right-chiral one
(Fig.~\ref{lptqdecay}).
Non-renormalizability here is provided with presence of an unremovable
non-pertubative process.

2) Left-chiral particles appear not in the same spatial point --
then summary spin can be compensated by orbital momentum of their relative
motion. Such interaction is described, in particular, by the following vertex:

\begin{equation}
L=\lambda_\parallel\bar q_{La}^i\gamma^\mu l_{Ll}\partial_\mu\chi_{al}^i
\label{my_beloved_vertex}
\end{equation}
In this paper we examine first possibility only. Note, however, that
(\ref {my_beloved_vertex}) must produce an unusual resonance cross-section
dependence upon the scattering angle, which, undoubtedly, of a considerable
importance for experimental theory verification. More than that, presence
of interactions, having non-local origin and similar to (\ref{my_beloved_vertex}),
seems to be a critical test for composite quark and lepton models.
This is the property which distinguishes the examined leptoquark from
those appearing in GUTs and other (preon-less) elementary particles theories.
We hope that this important property will help to determine wether DESY
discovery (if it is confirmed) can be considered as an argument for
quark and leptons inner structure existence.

Let's examine more properly the diagram shown at Fig.~\ref{lptqdecay}.
$VAC$ vertex represents effective sum of all local processes of leptoquark
decaying into left-chiral fermions pair (they are (\ref {preon_cond}) $U$- and
$D$-condensates interactions, pertubative birth of $U\bar U$($D\bar D$)-pair
from emitted by $x$-subpreon metagluon {\sl etc.}). After this process is
completed left-chiral quark transforms into right-chiral, which is necessary
as it has been mentioned above.

In fact, we have a solid process, a leptoquark interaction with correlated
condensates pair, as the first of described decay process components is itself
impossible. Nevertheless, the two formally introduced subprosesses arise
at a different energy hierarchy levels (preon and quark-lepton respectively),
and, consequently, we can suppose them to be independent in a high degree.
When this is accepted, the fact, that carrying flavour indices subpreons
doesn't take any part in $VAC$ interaction, makes us to think that relations
of different generation leptoquarks decays are determined only by the
diagram part, containing quark vacuum interaction. Thus, flavour hierarchy
of effective leptoquark coupling constants is similar to quark masses hierarchy.
We establish low-energy description of Fig.~\ref{lptqdecay} diagram
analogously to quark masses generation low-energy description
(i.e. substituting quark condensate interactions with interactions with
Higgs one):

\begin{equation}
L_{leptoquark}=\lambda_1m_{ab}\chi_{al}^i\bar q\rule{1pt}{0mm}{}_{Rb}^i(h_1^+l_{Ll})+
               \lambda_2\tilde m_{ab}\chi_{al}^i\bar{\tilde q}\rule{1pt}{0mm}{}_{Rb}^i
                 (h_2^+l_{Ll})
               +\mbox{h.c.}
\label{lptqrk_decay}
\end{equation}

Note, that alternatively to (\ref {lptqrk_decay}) we can in explicit form
write renormalizable vertices, corresponding to leptoquark decay, without
entering any Higgs fields into them. In this way we, however, would obtain
initially broken gauge symmetry Lagrangian. Within the above-introduced (in
our mind, more logical) approach, we arrive to a similar state after
spontaneous breaking.

\section{Leptoquark resonance influence on the cross-section of
$e^+d$-scattering}

\setcounter{equation}{0}
\renewcommand{\theequation}{3.\arabic{equation}}

In the beginning of 1997 year two independent research groups at
Hamburg collider HERA announced that the number of $e^+p\to e^+X$ and
$e^+p\to \bar\nu_eX$ events exceeds SM predictions (beams momenta were
$p_e=27.5GeV$ and $p_p=820GeV$). One of the suggested hypotheses explaining
such an anomaly is that kinematic variables of the experiment turned out to
be within the leptoquark resonance boundaries. Supposing that the suggested
leptoquark has preonic origin we can easily calculate its influences on
$e^+d\to e^+d$ and $e^+d\to u\bar\nu_e$ processes within the built in
section 2 effective low-energy $U(1)\times U(1)\times SU(2)$ theory.

After spontaneous symmetry breaking the part of interaction Lagrangian,
corresponding to neutral current process is written in the form:

\begin{equation}
\begin{array}{l}
\displaystyle L_{(e^+d\to e^+d)}=-\frac e3\bar d\hat Ad
                   -e\bar e\hat Ae
-\frac{\beta_d}6\bar d\hat Z(1+\delta_d\gamma^5)d
-\frac{\beta_e}2\bar e\hat Z(1+\delta_e\gamma^5)e-
\\
\rule{0pt}{8mm}
\rule{90pt}{0mm}
\displaystyle-\frac{\beta'_d}6\bar d\hat Z'(1+\delta'_d\gamma^5)d
-\frac{\beta'_e}2\bar e\hat Z'(1+\delta'_e\gamma^5)e+
\\
\rule{0pt}{8mm}
\rule{95pt}{0mm}
+G_{ed}\left[\chi_{11}^i\bar d^i(1-\gamma^5)e+
             \chi_{11}^{i+}\bar e(1+\gamma^5)d^i\right]
\end{array}
\label{neut_lagrange}
\end{equation}
To examine the charged current process we also need the following vertices:

\begin{equation}
\begin{array}{l}
\displaystyle L_{(e^+d\to u\bar\nu_e)}=
\frac{g_2}{2\sqrt{2}}\left(
\bar u^i\hat W^+d^i+\bar d^i\hat W^-u^i+\bar\nu_e\hat W^+e+\bar e\hat W^-\nu_e
\right)+
\\
\rule{0pt}{8mm}
\rule{75pt}{0mm}
+G_{u\bar\nu_e}\left[\chi_{11}^i\bar u^i(1-\gamma^5)\nu_e+
                     \chi_{11}^{i+}\bar\nu_e(1+\gamma^5)u^i\right]
\end{array}
\label{charged_lagrange}
\end{equation}
Effective broken symmetry Lagrangian parameters are expressed through gauge
constants, vacuum shifts and leptoquark coupling constants.
Thus $G_{ed}=\lambda_1\rho_1m_{11}/2$, $G_{u\bar\nu_e}=\lambda_2\rho_2\tilde m_{11}/2$.

When calculating the cross-sections we take into account a finite leptoquark
width within Breit-Wigner approximation. All below-written results are
obtained in chiral limit, which is quite logical when reactions energies
are about four orders of magnitude higher than the fermion masses.
In this approach leptoquark width (due to $e^+d$ and $u\bar\nu_e$ decay
processes) is equal to:

\begin{equation}
\Gamma=\frac1{4\pi}\left(G_{ed}^2+G_{u\bar\nu_e}^2\right)m_\chi
\label{lptqrk_width}
\end{equation}

The following four loopless diagrams contribute to elastic scattering:
transformation into leptoquark, succeeded with its reverse decay into
$e^+d$-pair ($s$-channel) and exchanges with photon, $Z$- and $Z'$-bosons in
$t$-channel. The corresponding expression in chiral limit in centre of mass
related frame of reference is:

\begin{equation}
\begin{array}{l}
d\sigma=\displaystyle\frac{\sin\chi d\chi}{64\pi}\left\{
\frac{32G_{ed}^4p^2}{(m_\chi^2-s)^2+m_\chi^2\Gamma^2}+
\frac83G_{ed}^2e^2\frac{m_\chi^2-s}{(m_\chi^2-s)^2+m_\chi^2\Gamma^2}\,
\frac1{\sin^2(\chi/2)}+
      \phantom{\frac{e^2}{\left(m_Z^2\right)}}
\right.
\\
\rule{0pt}{10mm}
\rule{25pt}{0mm}
+\displaystyle\frac83G_{ed}^2\beta_e\beta_d
\frac{m_\chi^2-s}{(m_\chi^2-s)^2+m_\chi^2\Gamma^2}\,
\frac{(1-\delta_e)(1+\delta_d)p^2}{m_Z^2+4p^2\sin^2(\chi/2)}+
\\
\rule{0pt}{10mm}
\rule{25pt}{0mm}
+\displaystyle\frac83G_{ed}^2\beta'_e\beta'_d
\frac{m_\chi^2-s}{(m_\chi^2-s)^2+m_\chi^2\Gamma^2}\,
\frac{(1-\delta'_e)(1+\delta'_d)p^2}{m_{Z'}^2+4p^2\sin^2(\chi/2)}+
\\
\rule{0pt}{10mm}
\rule{25pt}{0mm}
+\displaystyle\frac19\frac{e^4}{p^2\sin^4(\chi/2)}
\left(1+\cos^4(\chi/2)\right)+
\\
\rule{0pt}{10mm}
\rule{25pt}{0mm}
+\displaystyle\frac29\frac{\beta_e\beta_d\beta'_e\beta'_dp^2}
{\left(m_Z^2+4p^2\sin ^2(\chi/2)\right)
\left(m_{Z'}^2+4p^2\sin ^2(\chi/2)\right)}\times
\\
\rule{0pt}{6mm}
\rule{70pt}{0mm}
\times\left[
(1+\delta_e\delta'_e)(1+\delta_d\delta'_d)\left(1+\cos^4(\chi/2)\right)+
\right.
\\
\rule{0pt}{6mm}
\rule{70pt}{0mm}
\left.
+(\delta_e+\delta'_e)(\delta_d+\delta'_d)\sin^2(\chi/2)\left(\sin^2(\chi/2)
-2\right)\right]+
\\
\rule{0pt}{10mm}
\rule{25pt}{0mm}
+\displaystyle\frac19\frac{\beta_e^2\beta_d^2p^2}
{(m_Z^2+4p^2\sin ^2(\chi/2))^2}
\left[(1+\delta_e^2)(1+\delta_d^2)\left(1+\cos^4(\chi/2)\right)+
\right.
\\
\rule{0pt}{6mm}
\rule{70pt}{0mm}
\left.
+4\delta_e\delta_d\sin^2(\chi/2)\left(\sin^2(\chi/2)-2\right)\right]+
\\
\rule{0pt}{10mm}
\rule{25pt}{0mm}
+\displaystyle\frac19\frac{\beta_e^{'2}\beta_d^{'2}p^2}
{\left(m_{Z'}^2+4p^2\sin^2(\chi/2)\right)^2}
\left[(1+\delta_e^{'2})(1+\delta_d^{'2})\left(1+\cos^4(\chi/2)\right)+
\right.
\\
\rule{0pt}{6mm}
\rule{70pt}{0mm}
\left.
+4\delta'_e\delta'_d\sin^2(\chi/2)\left(\sin^2(\chi/2)-2\right)\right]+
\\
\rule{0pt}{10mm}
\rule{-13pt}{0mm}
+\displaystyle\frac29\frac{e^2\beta_e\beta_d}
{\sin^2(\chi/2)\left(m_Z^2+4p^2\sin^2(\chi/2)\right)}
\left[1+\cos^4(\chi/2)+
\delta_e\delta_d\sin ^2\frac\chi2\left(\sin^2\frac\chi2-2\right)\right]+
\\
\rule{0pt}{10mm}
\rule{-13pt}{0mm}
\left.
+\displaystyle\frac29\frac{e^2\beta'_e\beta'_d}
{\sin ^2(\chi/2)\left(m_{Z'}^2+4p^2\sin ^2(\chi/2)\right)}
\left[1+\cos^4(\chi/2)+
\delta'_e\delta'_d\sin ^2\frac\chi2\left(\sin ^2\frac\chi2-2\right)\right]
\right\}
\end{array}
\label{NC_cross}
\end{equation}
where $s=4p_ep_d$ is first Mandelstam invariant, $p=\sqrt{p_ep_d}$ --
scattering momentum, and $\chi$ -- scattering angle in center of mass related
frame of reference.
Scattering angle expression in laboratory frame of reference:

\begin{equation}
\theta=\arctan\left(
            \frac{2(p_ep_d)^{3/2}\sin\chi}
            {p_e^2p_d(1+\cos\chi)-p_ep_d^2(1-\cos\chi)}
            \right)
\label{theta}
\end{equation}
allows to calculate observable cross-section.

$e^+d\to u\bar\nu_e$ process is described by two diagrams only:
leptoquark one in $s$-channel and $W^+$-exchange graph in $t$-channel.
Cross-section expression structure in the case is quite similar to (\ref{NC_cross}):

\vspace{-5mm}
\begin{equation}
\begin{array}{l}
d\sigma=\displaystyle\frac{\sin\chi d\chi}{8\pi}p^2\left\{
\frac{4G_{ed}^2G_{u\bar\nu_e}^2}{(m_\chi^2-s)^2+m_\chi^2\Gamma^2}+
\displaystyle\frac14\frac{g_2^4}
{\left(m_{W^+}^2+4p^2\sin^2(\chi/2)\right)^2}\cos^4\frac{\chi}2
\right\}
\\
\rule{0pt}{8mm}
\rule{150pt}{0mm}
s=4p_up_{\bar\nu_e} \qquad p=\sqrt{p_up_{\bar\nu_e}}
\end{array}
\label{CC_cross}
\end{equation}

Note some important features of (\ref{NC_cross}) and (\ref{CC_cross}) expressions:

1) The cross-section has a sharp maximum in the vicinity of $p=m_\chi/2$ point
(so-called leptoquark resonance)

2) Interference effects brings to peak asymmetry in the neutral current process.
With some proper (\ref{neut_lagrange}) and (\ref{charged_lagrange}) Lagrangians
constants values notable holes in cross-section graph may appear.
Their location relatively to the peak is determined by positiveness or
negativeness of Lagrangian constants pair products.

3) In case of $e^+d\to u\bar\nu_e$ scattering $s$- and $t$-channels interference
is supressed due to the fact that $W^+$-boson interacts with left-chiral
quarks only. Cross-section peak then must be symmetric of a great degree.

4) For elementary processes (scattering of quarks, but not hadrons) the
peak height doesn't depend on leptoquark coupling constant -- the later
can influence it's width only.

5) The resonance is most easily observable in the case of backward scattering.

We have examined (\ref{NC_cross}) and (\ref{CC_cross}) expressions behavior
for constant values, which seem to be realistic ones.
The built in section 2 effective low-energy model contains some
parameters, which are not fixed by known experimental facts
($Z'$-boson mass, different Higgs fields vacuum shifts ratios).
Recalling the preon model, however, helps to draw some conclusions
on those undetermined quantities.

\begin{figure}[h]
\begin{picture}(0,0)
\put(10,-50){\epsfbox{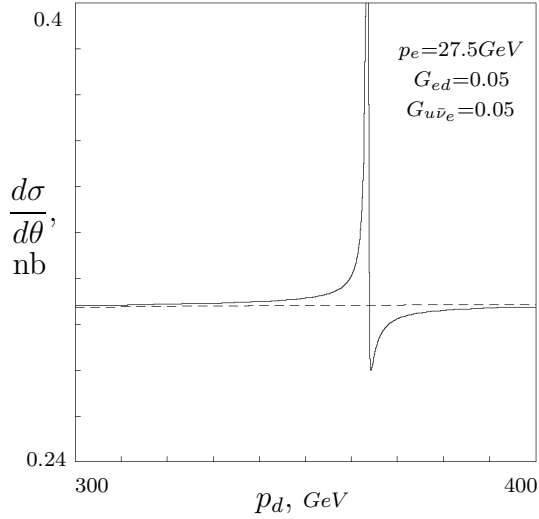}}
\put(1,-18){$\displaystyle \frac{d\sigma}{d\theta}$,}
\put(1.7,-25){nb}
\put(4.5,8){$\scriptstyle 0.4$}
\put(3.5,-50.5){$\scriptstyle 0.24$}
\put(34,-56){$p_d$, $\scriptstyle GeV$}
\put(10,-54){$\scriptstyle 300$}
\put(67,-54){$\scriptstyle 400$}
\put(53,4){$\scriptstyle p_e=27.5GeV$}
\put(55,0){$\scriptstyle G_{ed}=0.05$}
\put(54,-4){$\scriptstyle G_{u\bar\nu_e}=0.05$}
\end{picture}
\vspace{5.5cm}
\caption{Dependence graph of $e^+d\to e^+d$ elastic scattering to $\pi/4$
angle differential cross-section on $d$-quark beam momentum in laboratory frame
of reference.
Dashed line is SM cross-section.}
\label{ed_pi/4}
\end{figure}

\begin{figure}[h]
\begin{picture}(0,0)
\put(10,-50){\epsfbox{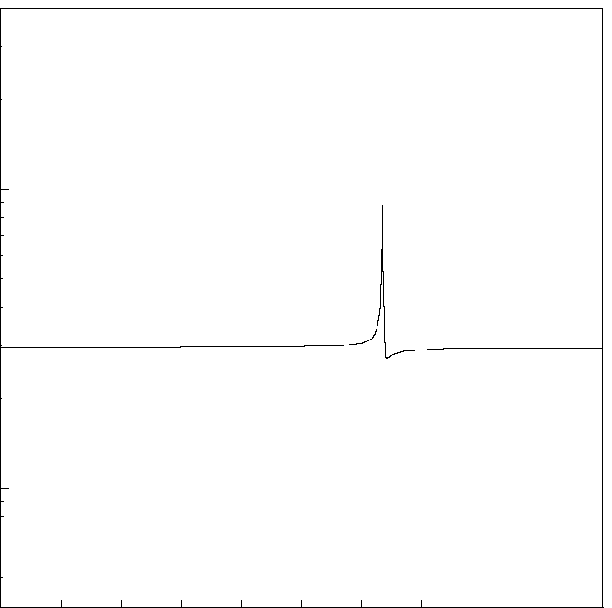}}
\put(1,-18){$\displaystyle \frac{d\sigma}{d\theta}$,}
\put(1.7,-25){nb}
\put(7,-9){$\scriptstyle 1$}
\put(5,-39.5){$\scriptstyle 0.1$}
\put(34,-56){$p_d$, $\scriptstyle GeV$}
\put(10,-54){$\scriptstyle 300$}
\put(67,-54){$\scriptstyle 400$}
\put(53,4){$\scriptstyle p_e=27.5GeV$}
\put(55,0){$\scriptstyle G_{ed}=0.05$}
\put(54,-4){$\scriptstyle G_{u\bar\nu_e}=0.05$}
\end{picture}
\vspace{5.5cm}
\caption{Dependence graph of $e^+d\to e^+d$ elastic scattering to $\pi/4$ angle
differential cross-section on $d$-quark beam momentum in laboratory frame of
reference (in logarithmic scale).}
\label{ln_ed_pi/4}
\end{figure}

\begin{figure}[h]
\begin{picture}(0,0)
\put(10,-50){\epsfbox{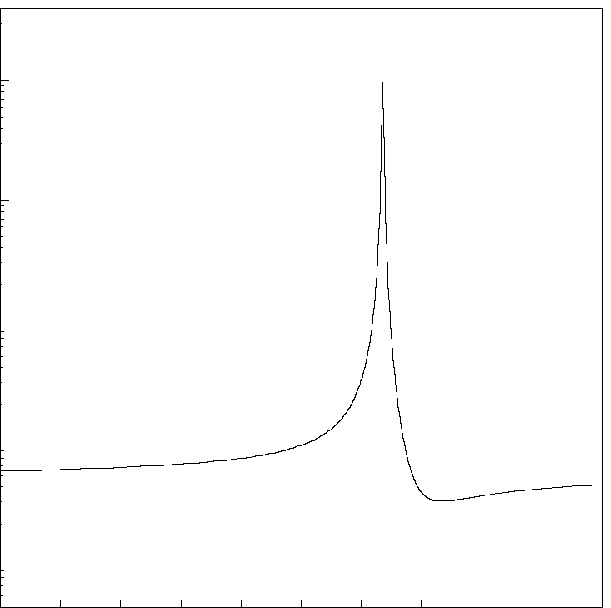}}
\put(1,-18){$\displaystyle \frac{d\sigma}{d\theta}$,}
\put(1.7,-25){pb}
\put(4.5,1.5){$\scriptstyle 10^{4}$}
\put(7,-47){$\scriptstyle 1$}
\put(34,-56){$p_d$, $\scriptstyle GeV$}
\put(10,-54){$\scriptstyle 300$}
\put(67,-54){$\scriptstyle 400$}
\put(53,4){$\scriptstyle p_e=27.5GeV$}
\put(55,0){$\scriptstyle G_{ed}=0.05$}
\put(54,-4){$\scriptstyle G_{u\bar\nu_e}=0.05$}
\end{picture}
\vspace{5.5cm}
\caption{Dependence graph of $e^+d\to e^+d$ elastic scattering to $3\pi/4$ angle
differential cross-section on $d$-quark beam momentum in laboratory frame of
reference (in logarithmic scale).}
\label{ln_ed_3pi4}
\end{figure}

\begin{figure}[h]
\begin{picture}(0,0)
\put(10,-50){\epsfbox{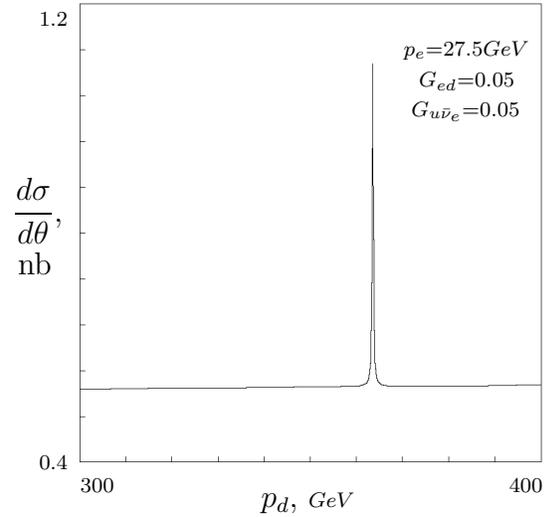}}
\put(1,-18){$\displaystyle \frac{d\sigma}{d\theta}$,}
\put(1.7,-25){nb}
\put(4.5,8){$\scriptstyle 1.2$}
\put(4.5,-51){$\scriptstyle 0.4$}
\put(34,-56){$p_d$, $\scriptstyle GeV$}
\put(10,-54){$\scriptstyle 300$}
\put(67,-54){$\scriptstyle 400$}
\put(53,4){$\scriptstyle p_e=27.5GeV$}
\put(55,0){$\scriptstyle G_{ed}=0.05$}
\put(54,-4){$\scriptstyle G_{u\bar\nu_e}=0.05$}
\end{picture}
\vspace{5.5cm}
\caption{Dependence graph of $e^+d\to u\bar\nu_e$ inelastic scattering to $\pi/4$
angle differential cross-section on $d$-quark beam momentum in laboratory frame
of reference.}
\label{unu_pi/4}
\end{figure}

As it has been said in section 2.1, vacuum shifts of $h_1$ and $h_2$ doublets
are close to each other, but can't be equal precisely, for this will cause
a zero axial neutral Higgs-boson mass. In $e^+d$-scattering calculations,
however, assuming sharp doublet shifts equality doesn't carry to any
anomal consequences. All physical quantities show a smooth behavior when
changing $\cos^2\varphi$ parameter, which vanishes when the doublets shifts are
equal. Thus, within the approximate calculation we can put $\rho_1=\rho_2$,
i.e. $\varphi=\pi/4$. $g_2$ and $g_W$ constants having similiar origin at preon level
will also be considered as equal.
Within this approach effective Lagrangian parameters are expressed through
experimentally known quantities: electron charge $e$ and Weinberg angle
$\theta_W$:

\vspace{-5mm}
\begin{equation}
\begin{array}{l}
\hspace{-3mm}
\displaystyle\beta_d=\frac e{\sin2\theta_W}(3-4\sin^2\theta_W) \qquad
\displaystyle\delta_d=\frac3{4\sin^2\theta_W-3} \qquad
\displaystyle\beta'_d=-\frac32\frac e{\sin\theta_W} \qquad
\delta'_d=-1
\\
\rule{0pt}{8mm}
\hspace{-3mm}
\displaystyle\beta_e=\frac e{\sin2\theta_W}(1-4\sin^2\theta_W) \qquad
\displaystyle\delta_e=\frac1{4\sin^2\theta_W-1} \qquad
\displaystyle\beta'_e=-\frac e{2\sin\theta_W} \qquad
\delta'_e=-1
\end{array}
\label{constants}
\end{equation}
For essential masses the following values were used: $m_\chi=200GeV$,
$m_{W^+}=82GeV$, $m_Z=91GeV$, $m_{Z'}=465GeV$. We studied $(d\sigma/d\theta)$
graphs with a fixed scattering angle and positron beam energy ($p_e=27.5GeV$)
and varied energy of the colliding quarks; $G_{ed}$ and $G_{u\bar\nu_e}$
constants were put equal to $0.05$. The results (Fig.~\ref{ed_pi/4}-\ref{unu_pi/4})
straightforwardly
confirm all the stated general properties of the resonance scattering process.

\twocolumn

\vphantom{top}
\vspace{1.5cm}

\vspace{2cm}

\newpage

\vphantom{top}
\vspace{1.5cm}

\vspace{1.89cm}

\onecolumn

\section{Low-energy empiric predictions status}

The obtained in previous section expressions (\ref{NC_cross}) and (\ref{CC_cross})
don't belong to the set of quantities which are directly experimentally
observable. Starting from them, however, we can easily come to inclusive
cross-sections of $e^+p\to e^+X$ and $e^+p\to \bar\nu_eX$ processes, which are
examined in DESY. Even now we can draw some conclusions on their properties.

The first striking discordance between graphs at Fig.~\ref{ed_pi/4}-\ref{unu_pi/4}
and German researchers data is that peak cross-section value (independently upon
leptoquark coupling constant) exceeds background one in a
few orders of magnitude, but H1 and ZEUS groups announced just about
observation of events, which number surpasses SM predictions about a time with
half only. Under a scrutiny this argument falls, for we examine only an
elementary process, but not an inclusive cross-section. In case of proton
scattering partons spread over momentum brings to a dependence of inclusive
cross-section peak value not only on elementary process peak cross-section
value, but also on resonance width, which itself depends on leptoquark
coupling constants values. Hence, leptoquark-fermion interactions weakness
can cause an unboundedly strong resonance slackening. All the rest,
said on quark-lepton cross-sections in the previous section can, probably,
without notable changes be considered as statements about deep-inelastic
scattering features.

We would like to stress, that even an absolute coincidence of the built
cross-sections with those which are to be carried out of the future experiments,
can't be considered as a confirmation of the preon structure stated in
the beginning of the paper. Leptoquark resonance existence is predicted
by a vast class of preon models and by some theories with structureless
fermions. Nevertheless, the obtained cross-sections possess some features
which can be considered as a ``signature'' of some sort of leptoquarks:
scalar particles with chiral coupling to fermions, i.e. exactly those
leptoquarks which are predicted by preon models.

Thus, statistics volume increasing at HERA will allow to judge wether preon
notions development is reasonable. Unless first empirical cross-sections
behavior observations reject some most simple and general predictions
of preon leptoquark models, the further studying with paying great attention
to non-local interactions (similar to the one described in section 2.2)
will allow to obtain a definitive answer about vital capacity of the
developed concepts.

Concluding, the author would like to address some warm thanks to G.M.Vereshkov,
without whom this work would be absolutely impossible.

\end{document}